# OPTIMAL RESONANT ASYMPTOTICS OF WAKEFIELD EXCITATION IN PLASMA BY NON-RESONANT SEQUENCE OF ELECTRON BUNCHES

*V.I. Maslov[1,2], E.O. Bilokon[2], V.O. Bilokon[2], I.P. Levchuk[1], I.N. Onishchenko[1]*
[1]*National Science Center "Kharkov Institute of Physics and Technology", Kharkiv, Ukraine;*
[2]*V.N. Karazin Kharkiv National University, Kharkiv, Ukraine*
*E-mail: vmaslov@kipt.kharkov.ua*

Resonant asymptotics of wakefield excitation in plasma by non-resonant sequence of relativistic electron bunches has been numerically simulated. It has been shown that in resonant asymptotics at optimal parameters the wakefield is excited with the maximum growth rate and the amplitude of the excited wakefield is the largest.

## INTRODUCTION

As plasma in experiment is inhomogeneous and nonstationary and properties of wakefield changes at increase of its amplitude it is difficult to excite wakefield resonantly by a long sequence of electron bunches (see [1, 2]), to focus sequence (see [3-7]), to prepare sequence from long beam (see [8-10]) and to provide large transformer ratio (see [11-17]). In [2] the mechanism has been found and in [18-21] investigated of resonant plasma wakefield excitation by a nonresonant sequence of short electron bunches. The frequency synchronization results by defocusing of those bunches which fall into a wrong phase with respect to the wave. In [6] it has been shown that nonresonant wakefield also can effectively focuses the bunches. Here results are presented on 2.5D numeral simulation by code LCODE [22] of resonant asymptotics of wakefield excitation in plasma by non-resonant sequence of relativistic electron bunches. Under resonant asymptotics we mean the excitation of the wakefield with the maximum growth rate, when the non-resonant sequence has already self-cleaned so that the interaction of the excited wakefield with the bunch electrons in the acceleration phases is negligible. Then the wakefield grows with steps.

## 1. MODEL OF NUMERICAL SIMULATION

We consider the wakefield excitation, when initially the plasma density $n_{0e}$ is a little smaller than resonant one $n_{0e} < n_{res} (=\omega_m^2 m_e/4\pi e^2)$ ($\omega_m$ is the repetition frequency of bunches). I.e. we consider the non-resonant case, when $\omega_m$ is a little larger the plasma frequency $\omega_m > \omega_p$. We consider the conditions at which the largest wakefield amplitude is excited and largest its growth rate is obtained.

We use the cylindrical coordinate system (r, z). Time $\tau$ is normalized on $\omega_{pe}^{-1}$, distance – on $c/\omega_{pe}$, density – on $n_{res}$, beam current $I_b$ – on $I_{cr}=\pi mc^3/4e$, fields – on $(4\pi n_{res} c^2 m_e)^{1/2}$.

Parameters are taken close to those of plasma wakefield experiments [23].

## 2. RESULTS OF NUMERICAL SIMULATION

We first consider the wakefield excitation in a plasma of length 1.67 m by first 28 bunches of finite length (at half-height) $\xi_b=0.2\lambda$ (Fig. 1). One can see that 11-th and 12-th bunches are similar to the first bunch, and they just lead to the next increase in the amplitude of the wakefield (Fig. 2). From Fig. 2 one can see that after point, where the maximum wakefield amplitude is reached (Fig. 3), it grows approximately by step. This means that the intermediate bunches are defocused and do not interact with the wakefield. At the same time, bunches, which excite the wakefield, at the point of reaching maximum amplitude have parameters close to those which they had at the point of their injection.

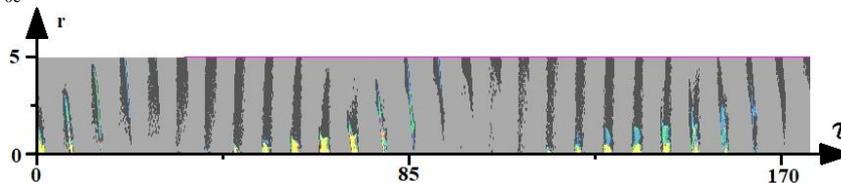

*Fig. 1. Temporal evolution of the beam density at $\gamma_b=5$; $(n_e-n_{res})/n_{res}=-0.15$, $\xi_b=0.2\lambda$, $I_b=0.526\cdot10^{-3}$, $z=1.33m$*

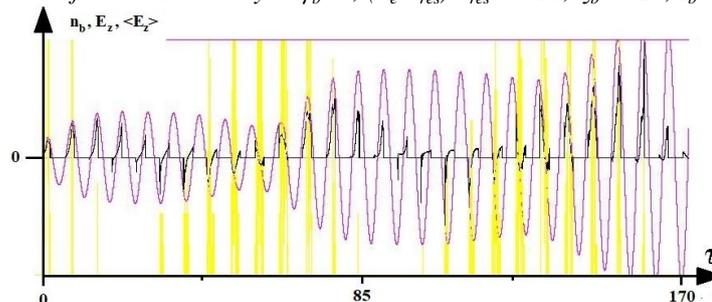

*Fig. 2. The on-axis wakefield excitation $E_z$ (red), $<E>=\int dr\, r\, E_z n_b/\int dr\, r n_b$ (black) and density of bunches on the axis $n_b=n_b(r=0)$ (yellow) as a function of the time $\tau$ for $\gamma_b=5$, $\xi_b=0.2\lambda$, $I_b=0.526\cdot10^{-3}$, $z=1.33m$ by train of 28 bunches in the nonresonant case $(n_e-n_{res})/n_{res}=-0.15$*

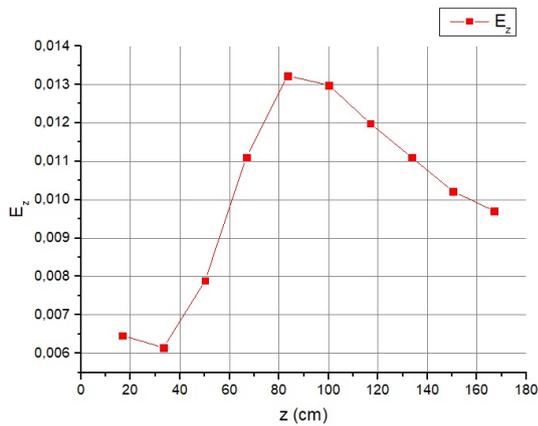

Fig. 3. The amplitude of $E_z$ as a function of the coordinate along the plasma at $\gamma_b=5$; $(n_e-n_{res})/n_{res}=-0.15$, $\xi_b=0.2\lambda$, $I_b=0.526\cdot 10^{-3}$

From Fig. 3 one can see that the maximum amplitude of excited wakefield is reached at a depth, equal z=83 cm. This depth corresponds to approximately 8 wavelengths. At this depth, the electrons of the bunches from the accelerating phases have been defocused such strongly that their interaction with wakefield can be neglected. At the same time, the radii of the decelerated bunches are approximately initial (Fig. 4).

In the vicinity of the injection point, both the heads of the first bunches at the beginning of each wakefield beating, and the tails of the last bunches at the end of each beginning are under the action of a radial force of approximately equal $F_r\approx 0$ (Figs. 5, 6). And the tails of the first bunches at the beginning of each beating and the head of the last bunches at the end of each beating are under the effect of a weak focusing force (see Figs. 5, 6).

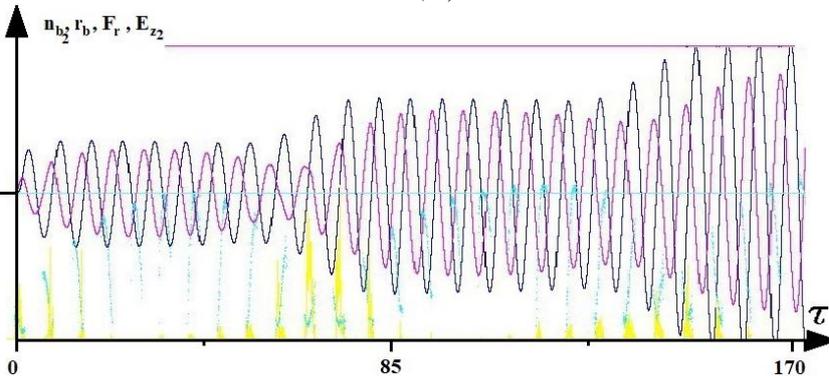

Fig. 4. $E_{z2}=E_z(r=r_b)$ (red), the radial wake force $F_r$ (dark blue), $n_{b2}=n_b(r=r_b)$ (yellow), and radius of bunches $r_b$ (blue) as a function of the time for $\gamma_b=5$; $(n_{0e}-n_{res})/n_{res}=-0.15$, $\xi_b=0.2\lambda$, $I_b=0.526\cdot 10^{-3}$, $z=1.33m$

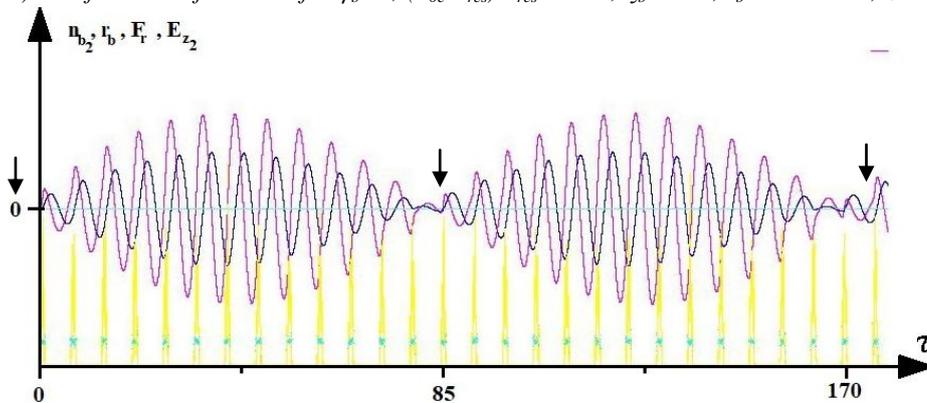

Fig. 5. $E_{z2}$ (red), $F_r$ (dark blue), $n_{b2}$ (yellow) and $r_b$ (blue) as a function of the time for $\gamma_b=5$; $(n_{0e}-n_{res})/n_{res}=-0.15$, $\xi_b=0.1\lambda$, $I_b=1.05\cdot 10^{-3}$, $z=0.05m$. 1-st bunches in the beginning of each beating are shown by arrows

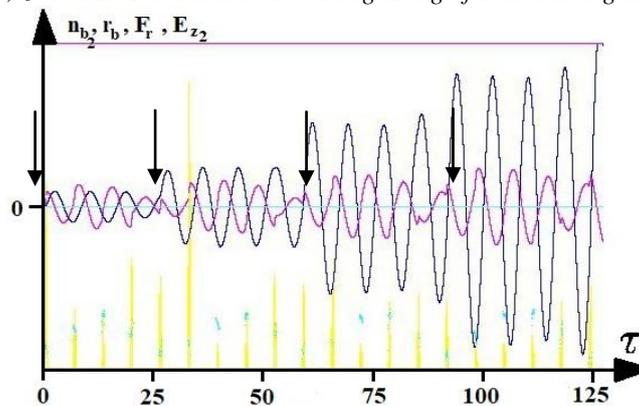

Fig. 6. $E_{z2}$ (red), $F_r$ (dark blue), $n_{b2}$ (yellow), and $r_b$ (blue) of $N_b=20$ bunches as a function of the time for $\gamma_b=5$; $(n_{0e}-n_{res})/n_{res}=-0.35$, $\xi_b=0.05\lambda$, $I_b=1.56\cdot 10^{-3}$, $z=1.0m$. 4 bunches, which excite wakefield, are shown by arrows

Taking into account that all beats at the beginning are the same and identical, the impression is that there is a symmetry between the behavior of the bunches at the beginning and at end of each beating. However, this symmetry is quickly violated, because with an increase of the wakefield amplitude, the beats are shortened. Thus, the bunches at the ends of the beats, which were in $F_r \approx 0$, get into the final $F_r$.

This gives an advantage to the first bunches at the beginning of each beating and the hope to find the optimum, taking into account that all bunches get a large defocusing impulse, and the 1st bunches at the beginning of each beating get a small focusing impulse. Optimal case corresponds to the case when the accelerated bunches are defocused as far as possible from the region of interaction with the wakefield $E_z$. In this case, the decelerating bunches does not have time to be significantly defocused, because they get into the phases of small $F_r$. And the remaining bunches get into the phases of large $F_r$. It can be assumed that the optimal case (the maximum amplitude of the wakefield and the maximum its growth rate) corresponds to the case when many bunches have already defocused, and those that lead to the growth of the wakefield were initially primarily focused, and then expanded to approximately the initial radius.

Since bunches at a large distance are expanded both in the defocusing and in the focusing fields, then the wakefield is excited effectively by the bunches which are under $F_r \approx 0$. If the bunches are very thin discs, then we can assume that the optimal case corresponds to the case when many bunches have already defocused, and those that lead to the stepwise growth of the wakefield have an initial radius, since they are in $F_r \approx 0$ (see Fig. 6).

It is also important, under what radial forces are the bunches when they move through the plasma. Those bunches that intensively excite the wakefield are not always in the focusing field during moving. They can move under the action of an alternating field, which varies in the vicinity of $F_r \approx 0$. I.e. they can first be under the action of a small focusing field $F_r > 0$, then in $F_r \approx 0$, then under the action of a small defocusing field $F_r < 0$.

There is a fundamental difference between resonant and nonresonant regimes. Namely, in the resonance case, only the 1st bunch is focused, while the remaining heads of bunches are defocused, and the tails of bunches are focused. In the nonresonant case $\omega_m > \omega_p$, all the bunches on the fronts of the beats are focused, and the remaining comparatively short bunches are strongly defocused.

## CONCLUSIONS

So, it has been shown by numerically simulated that in resonant asymptotics of wakefield excitation in plasma by non-resonant sequence of relativistic electron bunches at optimal parameters the wakefield is excited with the maximum growth rate and the amplitude of the excited wakefield is the largest. Up to resonant asymptotics the non-resonant sequence has already self-cleaned so that the interaction of the excited wakefield with the bunch electrons in the acceleration phases is negligible. Then the wakefield grows approximately with steps.

**ОПТИМАЛЬНАЯ РЕЗОНАНСНАЯ АСИМПТОТИКА ВОЗБУЖДЕНИЯ КИЛЬВАТЕРНОГО ПОЛЯ В ПЛАЗМЕ НЕРЕЗОНАНСНОЙ ПОСЛЕДОВАТЕЛЬНОСТЬЮ ЭЛЕКТРОННЫХ СГУСТКОВ**

*В.И. Маслов, Э.О. Билоконь, В.О. Билоконь, И.П. Левчук, И.Н. Онищенко*

Численно промоделирована резонансная асимптотика возбуждения в плазме кильватерного поля нерезонансной последовательностью релятивистских электронных сгустков. Показано, что в резонансной асимптотике при оптимальных параметрах кильватерное поле возбуждается с максимальным инкрементом, а амплитуда возбуждаемого кильватерного поля наибольшая.

**ОПТИМАЛЬНА РЕЗОНАНСНА АСИМПТОТИКА ЗБУДЖЕННЯ КІЛЬВАТЕРНОГО ПОЛЯ В ПЛАЗМІ НЕРЕЗОНАНСНОЮ ПОСЛІДОВНІСТЮ ЕЛЕКТРОННИХ ЗГУСТКІВ**

*В.І. Маслов, Е.О. Білоконь, В.О. Білоконь, І.П. Левчук, І.М. Оніщенко*

Чисельно промодельована резонансна асимптотика збудження в плазмі кільватерного поля нерезонансною послідовністю релятивістських електронних згустків. Показано, що в резонансній асимптотиці при оптимальних параметрах кільватерне поле збуджується з максимальним інкрементом, а амплітуда збуджуваного кільватерного поля найбільша.